\documentclass[aps,twocolumn,showpacs]{revtex4}%
\usepackage{amsfonts}
\usepackage{amsmath}
\usepackage{amssymb}
\usepackage{graphicx}

\begin{document}
\title{Vortices and ring dark solitons in nonlinear amplifying waveguides}
\author{Jie-Fang Zhang}
\address{Institute of
Nonlinear Physics, Zhejiang Normal University, Jinhua, Zhejiang
321004, China}
\author{Lei Wu}
\address{Tianmu College, Zhejiang Forestry University, Lin'an
311300, China}
\author{Lu Li}
\address{Institute of Theoretical Physics, Shanxi University,
Taiyuan, Shanxi 030006, China}

\author{Dumitru Mihalache}
\address{Horia Hulubei National Institute
for Physics and Nuclear Engineering, Magurele-Bucharest, 077125,
Romania}
\author{Boris A. Malomed}
\address{Department of Physical Electronics,
Faculty of Engineering, Tel Aviv University, Tel Aviv 69978,
Israel}

\begin{abstract}
We consider the generation and propagation of (2+1)-dimensional
beams in a nonlinear waveguide with the linear gain. Simple
self-similar evolution of the beams is achieved at the asymptotic
stage, if the input beams represent the fundamental mode. On the
contrary, if they carry vorticity or amplitude nodes (or phase
slips), vortex tori and ring dark solitons (RDSs) are generated,
featuring another type of the self-similar evolution, with an
exponentially shrinking vortex core or notch of the RDS. Numerical
and analytical considerations reveal that these self-similar
structures are robust entities in amplifying waveguides, being
\emph{stable} against azimuthal perturbations.
\end{abstract}

\pacs{42.81.Dp, 42.65.Tg, 05.45.Yv} \maketitle

%%%%%%%%%%%%%%%%%%%%%%%%%%%

The topic of self-similarity has been in the focus of intense
research interests in versatile areas of physics
\cite{selfsimilarity}. In nonlinear optics, the intrinsic
self-similarity of the nonlinear Schr\"{o}dinger (NLS) equation
has led to the development of the symmetry-reduction method, which
has been widely applied to search for exact and asymptotic
self-similar solutions
\cite{parabolic2,parabolic3,self_1,self_4,chang,Agrawal_2,serkin,Wise}.
Such optical waves (\textit{similaritons}) possess many attractive
features that make them potentially useful for various
applications to fiber-optic telecommunications and photonics, as
they can maintain the overall shape, while allowing their
amplitudes and widths to vary, following the modulation of
system's parameters -- the dispersion, nonlinearity, gain,
inhomogeneity, and others. Recent studies reveal that similaritons
also exist in other fields of physics, including Bose-Einstein
condensates and plasmas \cite{bec1,bec2,plasma2}.

Among various types of self-similar modes, there has been an increasing
interest in the study of asymptotically exact parabolic similaritons, since
their first experimental realization in normally dispersive fiber amplifiers
\cite{parabolic2}. One important advantage of these waves is that they are
free of collapse and filamentation at high power levels. Another remarkable
property of the similaritons is that the corresponding output profile is
completely determined by the input power, i.e., all initial profiles with the
same power evolve toward the same parabolic similariton
\cite{parabolic2,parabolic3,wabnitz}, which is very attractive for
applications such as the pulse (or beam) shaping or optical regeneration
\cite{phase_sharp,phase_sharp2}. This conclusion, however, is correct only
when the initial beams are of the fundamental type, i.e., they do not have
phase slips or phase singularities, and their amplitude profiles do not have
nodes. If the input beam carries a phase slip, a dark soliton in the
self-similar parabolic background is generated \cite{lei_oe}. On the other
hand, a spatiotemporal vortex torus will emerge if the input beam has an
embedded vorticity \cite{Mihalache_PRL,chen_prl,Greece}.

In this work we study the generation and propagation of optical beams inside
the two-dimensional nonlinear waveguide amplifier, focusing on effects of
initial phase singularities, phase slips, and amplitude nodes. The
nonlinearity of the medium is assumed to be \emph{self-defocusing}, with
refractive index $n=n_{0}-n_{2}I$, where $I$ is the beam's intensity, and
$n_{2}$ is positive (the cubic self-defocusing is possible in semiconductor
waveguides). The respective governing equation for the paraxial optical beam
in such an amplifying waveguide is the (2+1)-dimensional NLS equation,
\begin{equation}
i\frac{\partial{u}}{\partial{z}}+\frac{1}{2}\left(  \frac{\partial{^{2}}%
}{\partial{x^{2}}}+\frac{\partial{^{2}}}{\partial{y^{2}}}\right)
u-|u|^{2}u=i\frac{g}{2}u, \label{temporal}%
\end{equation}
where the beam's envelope $u$, propagation distance $z$, spatial coordinates
$x$ and $y$, and gain $g$ are normalized by $(k_{0}n_{2}L_{D})^{-1/2}$,
$L_{D}$, $w_{0}$, and $L_{D}^{-1}$, respectively, with $k_{0}=2\pi
n_{0}/\lambda$, $L_{D}=k_{0}w_{0}^{2}$, and $w_{0}$ being, respectively, the
wavenumber at the input wavelength $\lambda$, diffraction length, and a
characteristic transverse width.

Our first aim is to investigate the self-similar evolution of the
optical beam in the asymptotic limit of $z\rightarrow\infty$ by
considering the intrinsic self-similarity of Eq. (\ref{temporal}).
For the evolution to be exactly self-similar, the functional form
of the beam's intensity profile must remain unchanged at different
propagation distances. Accordingly, one can express the beam's
intensity as
$|u(z,x,y)|^{2}={\exp[\int_{0}^{z}g(z^{\prime})dz^{\prime
}]}|U(X,Y)|^{2}/{\ell}^{2}$, where $(X,Y)\equiv(x,y)/\ell$ with
$\ell(z)$ being a positive function of the propagation distance
that characterizes the evolution of the beam's width, and
$|U\left( X,Y\right)  |^{2}$ is a functional form satisfying the
energy conservation condition, obtained after the subtraction
of the gain effect:%
\begin{equation}
\int\int|U(X,Y)|^{2}dXdY=U_{0}, \label{constraints}%
\end{equation}
with $U_{0}$ being the input power of the optical beam. With this
intensity shape, the expansion velocity of the beam's intensity
profile is $\mathbf{v} =\ell^{-1}\left(  {d\ell/d}z\right)
\mathbf{r}$, where $\mathbf{r} \equiv\left\{  x,y\right\}$. This
velocity is equal to the gradient of the beam's phase,
$\mathbf{v}=\nabla\phi$ \cite{book}. Therefore, the phase of the
beam undergoing the self-similar evolution is $\phi=\left(
{2\ell}\right) ^{-1}\left( {d\ell/d}z\right)  r^{2}-\phi_{0}(z)$,
where $\phi_{0}(z)$ is a phase offset. Thus, by introducing the
following self-similar transformation,
\begin{equation}
u\left(  x,y,z\right)  =\frac{U\left(  X,Y\right)  }{\ell}\exp\left[  \frac
{1}{2}\int_{0}^{z}g(z^{\prime})dz^{\prime}+i\phi\right]  ,
\label{transformation}%
\end{equation}
we cast Eq. (\ref{temporal}) into the form of%
\begin{align}
&  |U|^{2}U-{\frac{1}{2}\exp[-\int_{0}^{z}g(z^{\prime})dz^{\prime}]}\nabla
^{2}U\nonumber\\
&  =\exp[-\int_{0}^{z}g(z^{\prime})dz^{\prime}]\left[  \frac{d\phi_{0}}%
{dz}\ell^{2}-\frac{1}{2}\frac{d^{2}\ell}{dz^{2}}\ell^{3}R^{2}\right]  U,
\label{reduction}%
\end{align}
where $\nabla^{2}={\partial^{2}}/{\partial X^{2}}+{\partial^{2}}/{\partial
Y^{2}}$ and $R^{2}=X^{2}+Y^{2}$.

Without the loss of generality, we let $\left(  d^{2}\ell/dz^{2}\right)
\ell^{3}=\exp[\int_{0}^{z}g(z^{\prime})dz^{\prime}]$, and $\left(  d\phi
_{0}/dz\right)  \ell^{2}=\mu\exp[\int_{0}^{z}g(z^{\prime})dz^{\prime}]$, where
$\mu$ is a positive constant to be determined. Note that the coefficient in
front of the diffraction term in Eq. (\ref{reduction}), $\exp[-\int_{0}%
^{z}g(z^{\prime})dz^{\prime}]/2$, vanishes at
$z\rightarrow\infty$. Therefore, if $U\left(X,Y\right)$ is smooth
with finite value of ${\nabla^{2}U}$, the diffraction term may be
neglected in the asymptotic regime. The above consideration shows
that the exact self-similar evolution of the optical beam can be
achieved in the asymptotic limit, under which Eq.
(\ref{reduction}) yields
\begin{equation}
|U|^{2}=\mu-\frac{R^{2}}{2}, \label{bec}%
\end{equation}
for $R^{2}\leq2\mu$, and $U=0$ otherwise, where the positive
constant $\mu=\sqrt{U_{0}/\pi}$ is determined by Eq.
(\ref{constraints}). Note that this profile is essentially the
same as produced by the Thomas-Fermi approximation for the
ground-state solution of the two-dimensional Bose-Einstein
condensates described by the Gross-Pitaevskii equation with the
isotropic parabolic potential and repulsive interatomic
interaction \cite{GPE}. Thus, the exact self-similar beams are
robust entities.

Hereafter we assume $g$ to be a constant, hence the variable characterizing
the change of the beam width is
\begin{equation}
\ell(z)=\sqrt{\frac{4}{g}}\exp\left(  \frac{gz}{4}\right)  , \label{variable}%
\end{equation}
and the phase offset is
\begin{equation}
\phi_{0}(z)=\frac{\mu}{2}[\exp(\frac{gz}{2})-1]. \label{offset}%
\end{equation}
From Eqs. (\ref{transformation}) and (\ref{variable}), it follows
that the amplitude and the half width of the exact self-similar
beam increase exponentially as $\sqrt{\mu g}\exp(gz/4)/2$ and
$\sqrt{8\mu/g}\exp(gz/4)$, respectively. As the optical beam
expands, a phase-front curvature (\textit{chirp}) develops. Using
the stationary-phase method \cite{stationary}, one can find that
the spatial spectrum of the optical beam is parabolic too. Note
that the asymptotically exact self-similar evolution of the
optical beam is predicated upon the assumption that $U$ is smooth,
so that the relative strength of the diffraction becomes
negligible when compared to that of the nonlinearity \cite{chang}.
This condition does not hold if the input beam carries vorticity,
or nodes in its amplitude profile, or phase slips. As said above,
the generation of vortex tori and ring dark solitons (RDSs) is
expected in that case.

Now we proceed to the investigation of the dynamics of a vortex torus of
topological charge $S$ in the asymptotic limit. As the local beam's intensity
vanishes at the center of the vortex, the relative strength of the diffraction
in the vortical core is much larger than that of the nonlinearity. The size of
the core range is determined by the coherence length, $L=S\sqrt{1/2|u|^{2}}$.
In the presence of the gain, the beam's amplitude increases exponentially,
therefore $L$ decreases at the same rate. Therefore, at the asymptotic stage,
the vortex becomes very narrow when compared to the whole beam, which makes it
reasonable to seek for a solution to Eq. (\ref{reduction}) in the asymptotic
limit by setting
\begin{equation}\label{vortex}
    U=\Psi(R)\exp(iS\theta),
\end{equation}
where $\theta$ is the azimuthal angle, and neglecting terms
containing gradients of $\Psi$. Then, we obtain an asymptotic
expression for the local intensity of the vortex torus,
\begin{equation}
|\Psi|^{2}=\mu_{\mathrm{vort}}-\frac{R^{2}}{2}-\frac{p}{R^{2}},
\label{reduction4}%
\end{equation}
where $p\equiv(S^{2}/2)\exp(-gz)$, and $\mu$ in Eq. (\ref{bec}) is
replaced by $\mu_{\mathrm{vort}}$ such that the phase offset keeps
the form of Eq. (\ref{offset}) with $\mu$ replaced by
$\mu_{\mathrm{vort}}$.

From Eq. (\ref{reduction4}), it follows that, in this approximation, the power
in the core of the vortex vanishes at $R^{2}<\mu_{\mathrm{vort}}-\sqrt
{\mu_{\mathrm{vort}}^{2}-2p}\approx p/\mu_{\mathrm{vort}}$ (in the exact
solution, the beam's intensity vanishes as $R^{|S|}$ at $R\rightarrow0$). On
the other hand, the last term in Eq. (\ref{reduction4}) may be neglected at
large $R$, hence the beam's intensity profile keeps the parabolic form outside
of the core. Note that the local beam's intensity of the vortex torus at large
distances is slightly greater than that of the fundamental soliton [Eq.
(\ref{bec})] for the same input power, since the optical field is removed in
the narrow core of the vortex. Therefore, the difference of $\mu
_{\mathrm{vort}}$, which is generated by the local nonlinearity, from $\mu$ is
almost negligible. Thus it can be concluded, from the above expressions, that
the radius of the vortex decreases exponentially, as
\begin{equation}
r=R\ell\simeq\sqrt{\frac{2}{g\mu}}S\exp(-\frac{gz}{4}), \label{R}%
\end{equation}
whereas the width and amplitude of the whole beam increase
exponentially, like those of the fundamental spatial soliton. This
analysis predicts peculiarities of the self-similar evolution of
the vortex torus, in comparison with the vortex-free beam.

It should be emphasized that now $U$ is not only a function of $X$
and $Y$, but also a function of $z$ [Eqs. (\ref{vortex}) and
(\ref{reduction4})]. The latter contradicts to the definition of
the function $U=U(X,Y)$ in the transformation [Eq.
(\ref{transformation})]. This contradiction, however, is
negligible: if $U$ is also a function of $z$, Eq.
(\ref{reduction}) will contain an additional term
$i\ell^2\exp[-\int_{0}^{z}g(z^{\prime})dz^{\prime}]U_z$. From Eq.
(\ref{reduction4}) it follows that this term can be approximated
by $2i\exp(-gz/2)\mu_{vort}^{-1/2}pR^{-2}$, which is about
$\exp(gz/2)$ times less than the last term in Eq.
(\ref{reduction4}), and hence it is negligible at the asymptotic
limit. The same argument also holds for the dark solitons embedded
in the parabolic background shown below.

Figure 1 displays the evolution of an initial Laguerre-Gaussian
beam of topological charge $S=1$ towards the vortex torus, whose
intensity and phase profiles at the asymptotic stage of the
evolution are found to be in good agreement with the predictions
based on Eqs. (\ref{transformation}), (\ref{variable}),
(\ref{vortex}) and (\ref{reduction4}).  Further numerical
simulations show that initial Laguerre-Gaussian beams with
topological charges up to $S=10$ evolve into the corresponding
vortex tori, indicating that the latter are robust entities in the
waveguide amplifiers.

\begin{figure}[t]
\centering
\includegraphics[width=7cm,height=5cm]{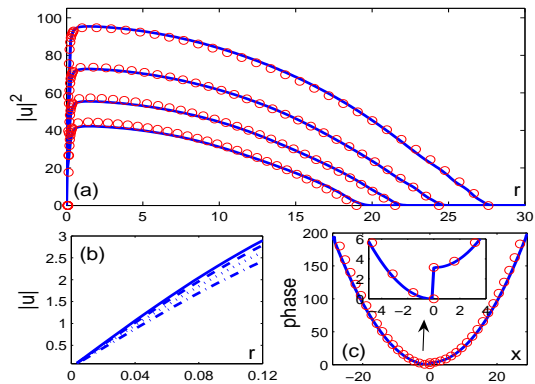}\caption{(a) The
self-similar evolution of the vortex torus of topological charge
$S=1$ in the waveguide amplifier, starting with input beam
$u(0,r)=r\exp(-{r^{2}}/{2} )\exp(i\theta)$, for $g=2$. From the
top to bottom, the propagation distance is $5.25$, $5$, $4.75$ and
$4.5$, respectively. (b) the beam's amplitude in the core of
vortex, which is a linear function of $r$. (c) The phase (phase
offset $\phi_0(z)$ is ignored) of the vortex torus at propagation
distance $z=5.25$, with the inset showing the phase slip $\pi$ in
the $x$ direction, to confirm that the topological charge is $1$.
Here and in following figures, the solid lines and circles
represent results of numerical simulations of Eq.
(\ref{temporal}) and the analytical prediction, respectively. }%
\end{figure}

Now, we proceed to the generation and propagation of dark solitons
in waveguide amplifiers. Since quasi-one-dimensional
(stripe-shaped) dark solitons eventually decay into vortices, we
focus on the RDS \cite{ring_dark}. Numerical simulations reveal
that dark solitons of this type can be generated from input beams
with nodes in the amplitude profile [Fig. 2(a)], or from Gaussian
beams with ring phase slips [Fig. 2(b)]. In the asymptotic limit,
the approximate RDS solution of Eq. (\ref{reduction}) can be
written as
\begin{equation}
|U|^{2}=\left(  \mu-\frac{r^{2}}{2\ell^{2}}\right)  \left\{  1-A^{2}%
\operatorname{sech}^{2}[\sqrt{\mu}A\exp\left(  \frac{gz}{2}\right)
\frac{r-r_{c}}{\ell}]\right\}  , \label{rds}%
\end{equation}
where $A$ and $r_{c}$ characterize the depth and location of the center of the
RDS, respectively. From Eq. (\ref{rds}) it follows that the effective width of
the black RDS decreases exponentially, as $\left(  2/\sqrt{g\mu}\right)
\exp(-gz/4)$, cf. Eq. (\ref{R}). The character of self-similar evolution of
the RDS, which is qualitatively similar to that of the vortex, is confirmed by
simulations displayed in Fig. 2, where one could also find that (i) the RDSs
generated from the initial phase slip are much more shallow than those
generated from the initial amplitude node, and (ii) the RDSs move outwards due
to their self-repulsive nature and the expanding parabolic background.
Finally, Fig. 3 shows the output beam intensity at $z=5$ in the amplifying
waveguide with $g=2$, as obtained from the input beam of power $\pi$, where
one could see that the vortex and RDSs may coexist, being embedded in the
self-similar parabolic background.

\begin{figure}[ptb]
\centering\includegraphics[width=7.5cm,height=3.5cm]{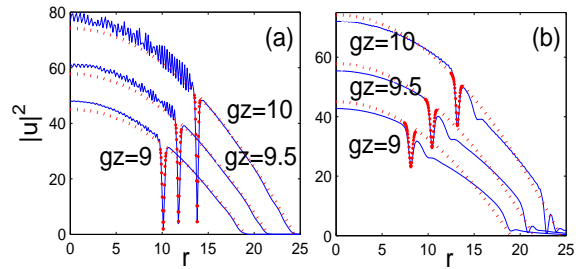}\caption{The
self-similar evolution of radial dark solitons in the amplifying
waveguide, obtained from input beams of total power $\pi$ with
initial profiles (a)
$u(0,r)=W^{-3}\exp(-{r^{2}}/{2W^{2}})(r^{2}-W^{2})$, and (b) $u(0,r)=W^{-1}%
\exp(-{r^{2}}/{2W^{2}})\exp\{i\pi\tanh[2(r-W)]/2\}$. The gain parameter is
$g=2$ and the initial width is $W=1.5$.}%
\end{figure}

\begin{figure}[ptb]
\centering
\includegraphics[width=7.5cm,height=5cm]{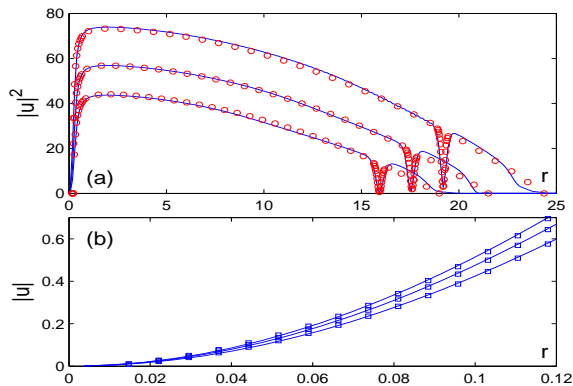}
\caption{(a) The coexistence of a vortex with the topological
charge $S=2$ and radial dark soliton, (b) the beam's amplitude in
the core of vortex (solid lines), where the squares represent the
numerical fits of $|u|$ as quadratic functions of $r$. The initial
condition is $u(0,r)=0.309\exp
(-r^{2}/2)(r^{2}-3/2)r^{2}\exp(2i\theta)$, and the gain parameter
is $g=2$.From the top to bottom, the propagation distance is $5$,
$4.75$ and $4.5$, respectively. }
\end{figure}

Next we investigate the stability of RDSs. By introducing the
transformation $U\rightarrow U\exp(-i\phi_{0})$, we rewrite Eq.
(\ref{reduction}) as
$iU_{Z}+\nabla^{2}U/[2\exp(gz)]-|U|^{2}U=R^{2}U/2$, where the new
propagation distance $Z\equiv(1/2)[\exp(gz/2)-1]$, as obtained
from Eq. (\ref{offset}). If the coefficient in front of the
diffraction term is $1$ ($g=0$),  this equation is just the
governing equation for the dynamics of RDS in Bose-Einstein
condensates with the harmonic potential \cite{ring_dark}, and it
was demonstrated that (i) in the limit case of a quasi-plane
soliton, the mass of the deep RDS is $2$ and its oscillation
frequency is $\omega=\sqrt{1/2}$, and (ii) the RDSs persists up to
$Z\simeq3T=6\pi/\omega$, and then begins to decay due to the
azimuthal modulation instability (AMI). When $g\neq0$, the
effective mass of the deep RDS is $2\exp(gz)$ and its oscillation
frequency is approximately $\sqrt{1/2}\exp(-gz/2)$. Thus, the RDS
would reach the AMI\ threshold at
$Z^{\prime}\approx6\sqrt{2}\pi\exp(gz/2)$. In fact, because $Z$ is
always smaller than $Z^{\prime}$, we infer that the instability
threshold will never be reached, in the present model, which is
indeed conformed by direct numerical simulations. Thus, the RDSs
are robust entities in the waveguide amplifiers, contrary to the
common opinion that AMI is inevitable for two-dimensional
topological modes, such as RDS
\cite{Mihalache_PRL,azimuthal,azimuthal2}.

In summary, within the framework of the (2+1)-dimensional NLS equation
including the linear gain, we have found three types of self-similar optical
beams in nonlinear amplifying waveguides, namely, the fundamental beams with
the parabolic profiles, robust vortex tori with different values of the
topological charge, and radial dark solitons with a parabolic background.
These results were obtained in the analytical form using the Thomas-Fermi type
of the asymptotic approximation, and confirmed by numerical simulations.

This work has been supported by the NNSF of China (Grant No.
10672147) and the Program for Innovative Research Team at Zhejiang
Normal University. L. Wu acknowledges the financial support from
Zhejiang Forestry University. The corresponding author is Jie-Fang
Zhang, his email address is jf\_zhang@zjnu.cn.

\end{document}